\documentclass{aa}

\usepackage{times}
\usepackage{graphics}

\begin{document}


\thesaurus{03
(11.09.3;
11.17.1;
11.17.4 HS 1603+3820)}


\title{HS\,1603+3820: a bright $z_{\rm em}=2.51$ quasar with a very
rich heavy element absorption spectrum}

\author{%
A. Dobrzycki
\inst{1}
\and
D. Engels
\inst{2}
\and
H.-J. Hagen
\inst{2}
}

\offprints{A. Dobrzycki}

\institute{Harvard-Smithsonian Center for Astrophysics, 60 Garden
Street, Cambridge, MA 02138, USA\\
e-mail: adobrzycki@cfa.harvard.edu
\and
Hamburger Sternwarte, Gojenbergsweg 112, D-21029 Hamburg, Germany\\
e-mail: [dengels,hhagen]@hs.uni-hamburg.de}

\date{Received / accepted}

\titlerunning{HS\,1603+3820: a rich complex of C{\sc iv} absorbers}

\maketitle


\begin{abstract}

During the course of Hamburg/CfA Bright Quasar Survey we discovered a
bright ($B=15.9$), high redshift ($z_{\rm em}=2.51$) quasar
HS\,1603+3820. The quasar has a rich complex of C{\sc iv} absorbers,
containing at least five systems, all within 3000~km\,s$^{-1}$ from
one another. Despite large ejection velocity ($v_{\rm
ej}>5000$~km\,s$^{-1}$ for all components) the complex is likely to be
associated with the quasar. There are at least three more associated
heavy element absorbers, two of which have $z_{\rm abs}>z_{\rm
em}$. Together, they make one of the richest known complexes of
associated heavy element absorbers, and the richest in objects with
$B<16$. The combination of redshift, brightness, richness of the metal
absorption spectrum and richness of the associated absorption is
unmatched among known quasars.

\keywords{intergalactic medium -- quasars: absorption lines --
quasars: individual: HS\,1603+3820}

\end{abstract}


\section{Introduction}\label{sec:intro}


\begin{figure}
\resizebox{\hsize}{!}{\includegraphics{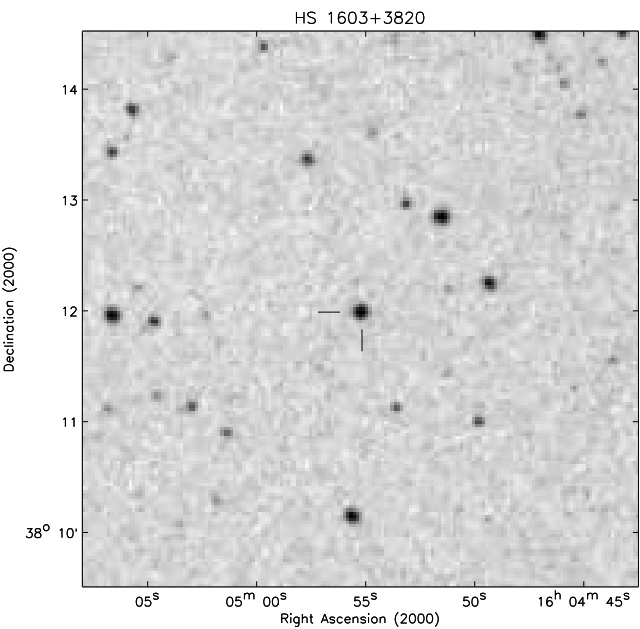}}
\caption{The finding chart for HS\,1603+3820.}
\label{fig:findingchart}
\end{figure}


\begin{figure}
\resizebox{\hsize}{!}{\includegraphics{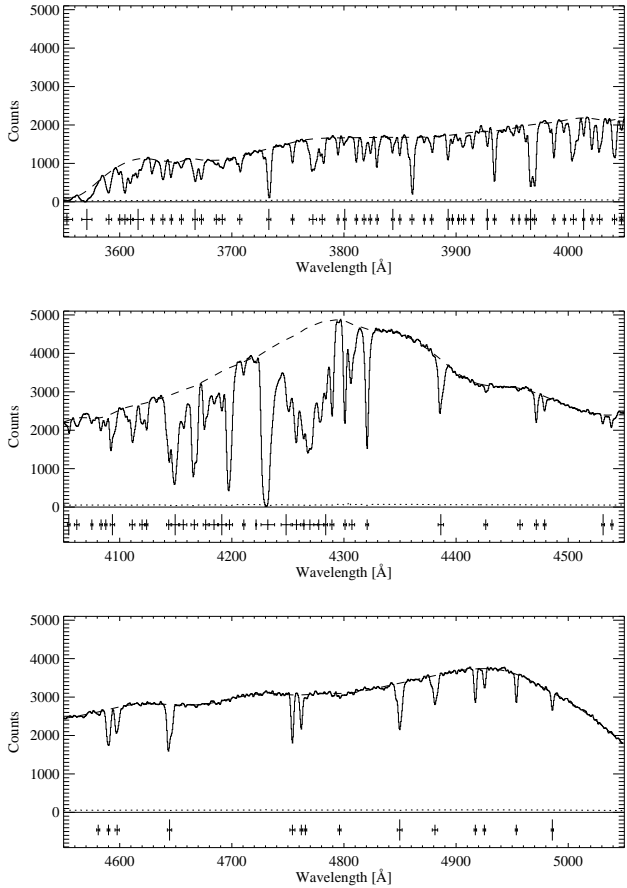}}
\caption{The ``Ly-$\alpha$'' spectrum of HS\,1603+3820. The dotted
line shows 1$\sigma$ uncertainty, the dashed line shows the continuum
it. Tick marks under the spectrum show the absorption lines. The
spectrum was smoothed with a 3-pixel box.}
\label{fig:lyaspectrum}
\end{figure}


\begin{figure}
\resizebox{\hsize}{!}{\includegraphics{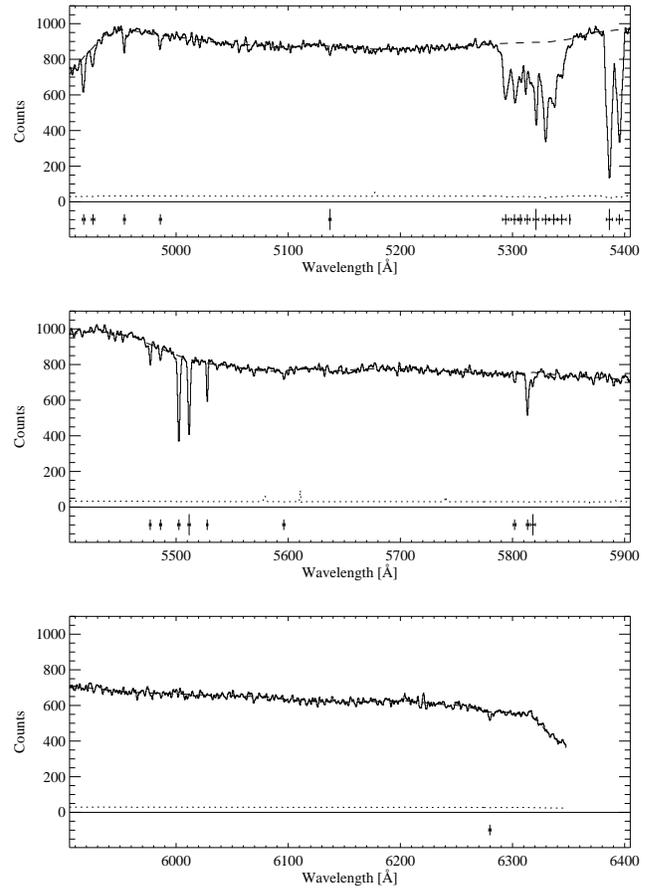}}
\caption{The ``C{\sc iv}'' spectrum of HS\,1603+3820. See caption of
Fig.~\ref{fig:lyaspectrum} for details.}
\label{fig:civspectrum}
\end{figure}


Quasars often exhibit narrow heavy element absorption lines near their
emission redshift. One possible explanation for the origin of these
systems is that they arise in clouds of matter associated with
galaxies in the clusters surrounding the quasars. Alternate
possibility is that they originate in the clouds that are physically
associated with the quasars themselves. It has been shown that both
these scenarios may be true (e.g.\ Ellingson et al.\ \cite{ell1994};
Hamann et al.\ \cite{ham1997a}).

Clusters of absorbers of any type are of particular interest. On one
hand, clustering properties of intervening systems depend strongly on
the type of absorbers, while on the other hand complexes of associated
absorbers give a unique opportunity to analyze the properties of their
hosts and of the quasar emission.

The stumbling block in such studies is the fact that bright high
redshift quasars with rich absorption are rare (for examples of such
systems see, e.g., Morris et al.\ \cite{mor1986}; Foltz et al.\
\cite{fol1987}; Petitjean et al.\ \cite{pet1994}; Hamann et al.\
\cite{ham1997a}; Lespine \& Petitjean \cite{les1997}; Petitjean \&
Srianand \cite{pet1999}). There are only a few quasars known with more
than just a handful of associated absorption systems which are bright
enough to perform high resolution spectral analysis.

In this paper we present a bright $z_{\rm em}=2.51$ quasar
HS\,1603+3820 ($\alpha_{\rm 1950} = \mbox{16:03:07.7}, \delta_{\rm
1950} = \mbox{+38:20:07}$) with a very rich metal absorption
spectrum. It was discovered during the course of the Hamburg/CfA
Bright Quasar Survey (Hagen et al.\ \cite{hag1995}; Dobrzycki et al.\
\cite{dob1996}).  Quasar candidates in the survey are selected from
the objective prism spectra taken with the Hamburg Schmidt telescope
at Calar Alto, Spain.  Follow-up low resolution spectroscopy is
performed with the 1.5-m Tillinghast reflector with the FAST
spectrograph at Fred Lawrence Whipple Observatory on Mount Hopkins,
Arizona.

The discovery spectrum and other basic information on HS\,1603+3820
will be presented in the forthcoming survey paper (Engels et al.\
\cite{eng1999}). In this paper we present a more detailed analysis of
the MMT spectrum of HS\,1603+3820. Its very unusual properties
prompted us to make the quasar available to interested researchers
prior to the publication of the survey.

To our knowledge, there are no radio or X-ray sources near the
position of HS\,1603+3820.

The finding chart for HS\,1603+3820 is shown on
Fig.~\ref{fig:findingchart}.  HS\,1603+3820 was selected for followup
studies because of its unusually high brightness ($B=15.9$) for a
high-$z$ quasar and because the discovery spectrum hinted that there
could be absorption systems in the vicinity of the emission lines.

Observations performed with the Multiple Mirror Telescope\footnote{MMT
is a joint facility of the Smithsonian Institution and the University
of Arizona.} revealed a very rich heavy element absorption spectrum.
Several metal systems, both intervening and associated, are present,
including a couple of systems with $z_{\rm abs}>z_{\rm em}$. A unique
feature is a rich complex of at least five C{\sc iv} absorbers near
the emission redshift of the quasar.

This paper is organized as follows.  In Sect.~\ref{sec:observations}
we present the MMT observations of HS\,1603+3820. In
Sect.~\ref{sec:metals} we present the metal absorption systems. In
Sect.~\ref{sec:complex} we discuss the of C{\sc iv} absorbers at
$z_{\rm abs}\approx z_{\rm em}$ and present arguments that they likely
are associated with the quasar. We summarize our results in
Sect.~\ref{sec:summary}.


\section{Observations}\label{sec:observations}

Spectra presented here were obtained during two nights, April 12-13,
1997, at the Multiple Mirror Telescope (MMT) on Mt.~Hopkins, Arizona.
We used the Blue Channel spectrograph, with the 1200~l/mm grating in
first order, 1.25$\times$3~arcsec slit and the 3k$\times$1k Loral CCD,
binned by two in the spatial direction. Weather and seeing were very
good during both nights. The wavelength range of 3530--5050~\AA, which
contains the Ly-$\alpha$ forest part of the HS\,1603+3820 spectrum,
was observed for the total of 3300~seconds during two nights. On the
second night another 1200~second exposure was obtained for the
wavelength range containing the C{\sc iv} emission line,
4840--6350~\AA.  Spectra have 0.495~\AA/pixel binning and spectral
resolution of $\sim$3~pixels. The S/N ratio per resolution element in
the ``Ly-$\alpha$ spectrum'' varies from $\sim$40 to $\sim$100, and in
the ``C{\sc iv} spectrum'' it is roughly uniform at the level of
45--50.


\begin{figure*}
\resizebox{\hsize}{!}{\includegraphics{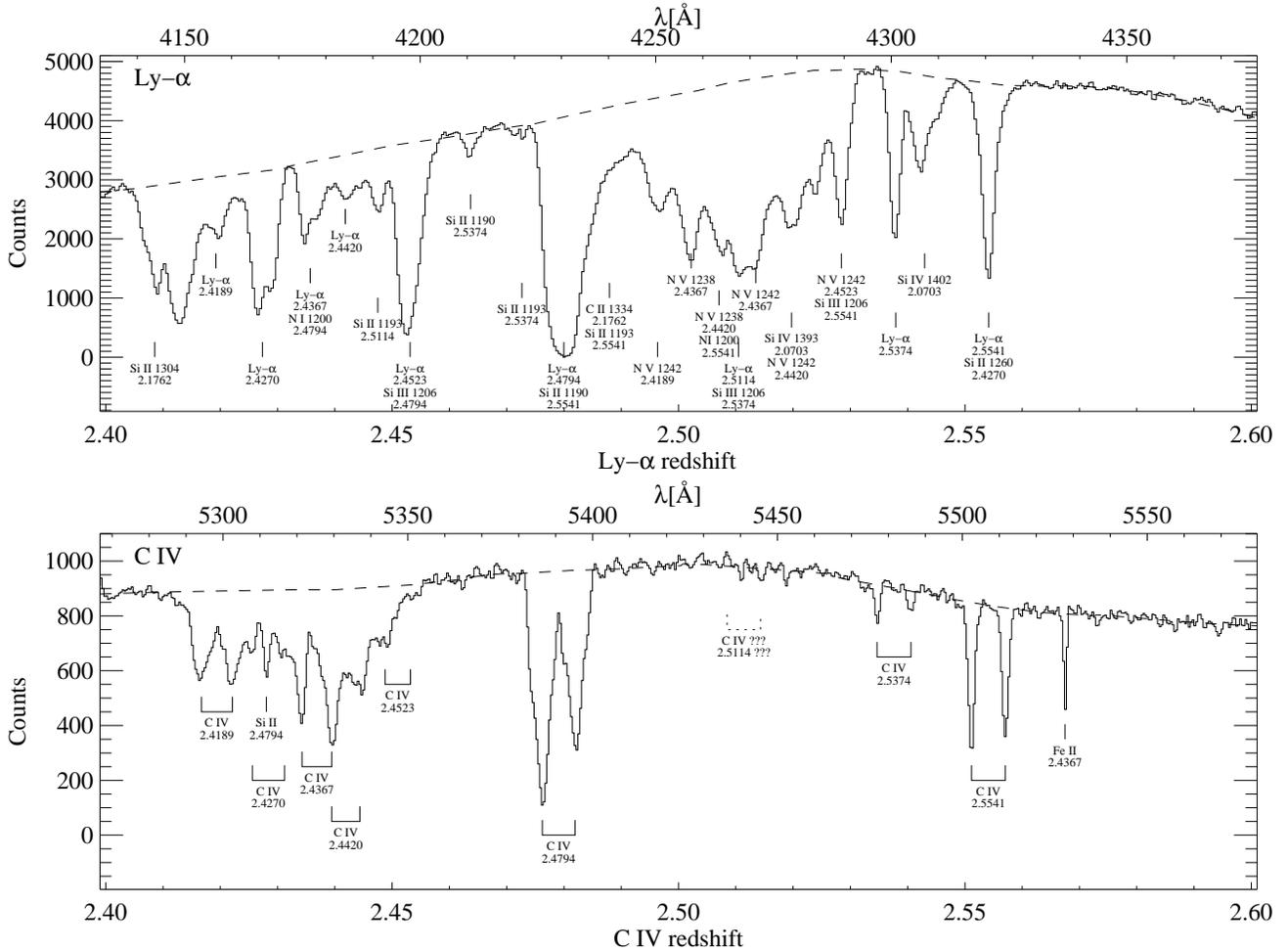}}
\caption{The vicinity of Ly-$\alpha$ (top panel) and C{\sc iv} (bottom
panel) emission lines, plotted in the same redshift range. Dashed line
shows the adopted continuum. Lines from identified metal systems are
marked. Note: the complex of C{\sc iv} absorbers at $z_{\rm
abs}=2.41-2.45$, very strong system at $z_{\rm abs}=2.4794$, two
systems on the red wings of the emission lines at $z_{\rm abs}=2.5374$
and $z_{\rm abs}=2.5541$, and missing C{\sc iv} absorption for the
uncertain system at $z_{\rm abs}=2.5114$ (its expected location is
marked with dotted line). See text for discussion.}
\label{fig:emlines}
\end{figure*}


Continua were fitted to the spectra and the absorption line lists were
generated following the procedure similar to outlined in Scott et al.\
(\cite{sco1999a}, \cite{sco1999b}), where one can also find the
analysis of the Ly-$\alpha$ forest spectrum of HS\,1603+3820.
Interested researchers can access the line lists (as well as finding
chart and the spectra -- both plots and digital versions) on the World
Wide Web at http://hea-www.harvard.edu/QEDT/Papers/hs1603/.

Full spectra can be seen in Figs.~\ref{fig:lyaspectrum}
and~\ref{fig:civspectrum}. Fig.~\ref{fig:emlines} shows in detail the
vicinity of the Ly-$\alpha$ and C{\sc iv} emission lines.

It has to be noted that the continuum level in the region near
4250~\AA, which occurs near the peak of the Ly-$\alpha$ emission line,
is quite uncertain, making the emission redshift of the quasar
somewhat uncertain, since an important part of the line is seriously
affected by complex absorption. (Please note that this absorption is
{\em not} related to Broad Absorption Line phenomenon -- the
absorption lines affecting the Ly-$\alpha$ emission profile are not
Ly-$\alpha$.)  We adopted the emission redshift of the quasar of 2.51,
keeping in mind the well known fact that broad emission lines of
quasars can be blueshifted by as much as $\sim$1000~km\,s$^{-1}$ with
respect to ``true'' quasar redshift, as determined from narrow
forbidden lines (see e.g.\ Espey \cite{esp1993}). It should be noted
(and it will be discussed in a little more detail below) that there
are absorption lines at redshift as high as 2.55, but they clearly are
on the red wing of the emission lines, well past the peak emission.


\section{Heavy element absorption systems}\label{sec:metals}

As expected in a quasar with $z_{\rm em}=2.51$, there are numerous
absorption features in the Ly-$\alpha$ forest part of the
spectrum. The Ly-$\alpha$ forest of HS\,1603+3820 has been included in
the large sample of Scott et al.\ (\cite{sco1999b}). In the present
paper we concentrate on the heavy element absorption lines, presence
of which is clearly seen on top of the emission lines and redwards of
the Ly-$\alpha$ emission.

Metal systems in the spectrum were searched for in two ways. First, an
attempt was made to interactively identify all absorption lines
redwards of the Ly-$\alpha$ emission line. Second, heavy element line
searching code METALS, written by Jill Bechtold and kindly provided to
us by the author, was applied. In all, thirteen heavy element
absorption systems were found, though the reality of two of them is
somewhat shaky. Notes on individual systems follow.

$\langle z_{\rm abs} \rangle = 1.8882$: This is the first of the two
uncertain systems. Its identification relies primarily on the presence
of a doublet which has the wavelength ratio matching the ratio for the
C{\sc iv} doublet. Though this match is indeed very good, both lines
are possible blends with lines from other systems. Two other
identified lines, N{\sc v}(1242) and Fe{\sc ii}(1608), are also blends
with lines from other systems. The wavelength corresponding to the
Ly-$\alpha$ line lies outside of the spectrum.

$\langle z_{\rm abs} \rangle = 1.9650, 2.0703, 2.1762$: All these
systems are unambiguously identified, showing well resolved
Ly-$\alpha$, C{\sc iv}, and other lines.

$\langle z_{\rm abs} \rangle = 2.4189, 2.4270, 2.4367, 2.4420,
2.4523$: These are the confirmed systems that have their C{\sc iv}
lines in the complex in the blended region near 5320~\AA. See
Sect.~\ref{sec:complex} below for a more detailed discussion of the
complex.

$\langle z_{\rm abs} \rangle = 2.4794$: This is a very strong system,
just below the emission redshift of the quasar. As many as seventeen
absorption lines belonging to this system were identified. The Si{\sc
ii}(1526) line lies inside the C{\sc iv} complex near 5320~\AA, but is
easily recognizable. Very strong Ly-$\alpha$ and C{\sc iv} lines are
prominent on Fig.~\ref{fig:emlines}.  The Ly-$\alpha$ absorption line,
which is somewhat asymmetric on the red wing, is in fact a blend of
(at least) two lines: very strong Ly-$\alpha$ proper, and a weaker
line, which appears to be a member of another metal system(s).

$\langle z_{\rm abs} \rangle = 2.5114$: This is the second of the two
uncertain systems. If it is real, it is almost exactly a $z_{\rm
abs}=z_{\rm em}$ system. It has six lines at correct wavelengths,
including three Si{\sc ii} lines. The Ly-$\alpha$ line lies in the
middle of the blended region near 4250~\AA. There is, however, no
significant C{\sc iv} absorption. Dotted marker on the lower panel of
Fig.~\ref{fig:emlines} shows the expected location of the C{\sc iv}
doublet; the 5$\sigma$ rest equivalent width threshold at this
location is 0.07~\AA. Also, there are two other issues that make the
identification of this system uncertain. First, the system appears to
contain a somewhat unusual combination of lines. Second, the ratios of
line strengths of the Si{\sc ii} lines appear to be incorrect (Morton
\cite{mor1991}). Acknowledging that especially the second of these
problems casts doubts on the reality of this system, we hesitate,
however, to entirely discard it, since the identified Si{\sc ii} lines
may be blended with other lines, not to mention the fact that the
environment in the vicinity of the quasar central engine is likely to
be highly unusual.

$\langle z_{\rm abs} \rangle = 2.5374,2.5541$: These are the two
systems with $z_{\rm abs}>z_{\rm em}$ seen in the spectrum of
HS\,1603+3820. Both systems have Ly-$\alpha$, C{\sc iv}, and other
metal lines clearly resolved. See below for discussion.


\section{Complex of metal systems at $z_{\rm abs} \approx
z_{\rm em}$: intervening or associated?}\label{sec:complex}

The most interesting feature in the spectrum of HS\,1603+3820 is the
large number of systems with $z_{\rm abs} \approx z_{\rm em}$. Nine of
the metal systems are within 8000~km\,s$^{-1}$ of the quasar emission
redshift.  Of these, the most spectacular is a complex of at least
five C{\sc iv} absorption systems, all within 3000~km\,s$^{-1}$ from
one another. Other systems may be present in the complex, but limited
resolution of our data does not allow us to claim that at a reasonable
level of confidence.

The average ejection velocity of the complex, i.e.\ displacement from
the redshift of the quasar in velocity units, is
$\sim$6500~km\,s$^{-1}$. In principle, large ejection velocities would
indicate that these systems are intervening, i.e.\ are not associated
physically with the quasar. On the other hand, convincing arguments
were presented (Hamann et al.\ \cite{ham1997b}, Richards et al.\
\cite{ric1999}) that absorbers with much larger ejection velocities
could be intrinsic to the quasar.

Of the three methods for distinguishing between intrinsic and
intervening systems (see Barlow \& Sargent \cite{bar1997}, Hamann et
al.\ \cite{ham1997a}, and references therein) -- details of shapes of
absorption lines, temporal variability of line strengths, and partial
covering of quasars by the absorbers -- only the third can be applied
in our case. The first is precluded by insufficient resolution of our
data. The second was made impossible by decommissioning of the MMT,
since we could not observe the quasar again with the same
telescope/detector combination, which is a standard approach in such a
case. We can, however, attempt to estimate whether at least one of the
systems in the complex obscures the emission source in the quasar
entirely. If it does not, it is a strong indication that the cloud in
which the absorption system originates is physically associated with
the quasar.

The C{\sc iv} doublet at $\langle z_{\rm abs} \rangle = 2.4189$, which
has $v_{\rm ej}\approx8000$~km\,s$^{-1}$, is at the high ejection
velocity end of the C{\sc iv} complex. The lines from this doublet
fulfill all the conditions needed for the quasar covering factor
criterion: they are satisfactorily well resolved, they are markedly
broader than the spectrum resolution, and they are not contaminated by
absorption lines from any other known systems. If these two lines are
indeed resolved, we find the C{\sc iv}(1548)/C{\sc iv}(1550) ratio in
this system to be close to 1, which indicates that the system is
optically thick. At the same time, the lines do not reach zero
intensity in their bottoms, which suggests that the cloud in which the
lines originate does not obscure the entire continuum emitting
source. That in turn suggests that this absorption system is intrinsic
to the quasar. The lower ejection velocity systems in the complex also
do not reach zero intensity, which makes it reasonable to assume that
they are also associated with the quasar (even though we cannot
unambiguously establish whether they are saturated).

There are three (or four, if one includes the uncertain $z_{\rm
abs}=z_{\rm em}$ system) other heavy element systems with ejection
velocities lower than the complex. Two of these systems are clearly
infalling: even if the emission redshift of the quasar is indeed
underestimated by as much as 1000~km\,s$^{-1}$, these systems still
have $z_{\rm abs}>z_{\rm em}$; their infall velocities are $\sim$2000
and $\sim$3600~km\,s$^{-1}$ with respect to adopted emission redshift
of the quasar.

Arguments presented above suggest that all eight (or nine) systems are
physically associated with the quasar. If this is the case, then all
systems need to be considered as one large associated absorption
complex, with dispersion of $\sim$4300~km\,s$^{-1}$.

It appears that the absorption originates in clouds in the immediate
vicinity of the quasar. The ejection velocities and the velocity
dispersion of the complex (both of the order of thousands of
km\,s$^{-1}$) appear to be too high to interpret the absorbers as
originating in the halos of galaxies in the quasar host cluster.


\section{Summary}\label{sec:summary}

The combination of high redshift, brightness, richness of the heavy
element absorption -- and associated absorption in particular -- in
the spectrum of HS\,1603+3820 is truly unique. Eleven confirmed C{\sc
iv} systems in the spectrum of HS\,1603+3820 ranks among the richest
known. The York catalog (York et al.\ \cite{yor1991}, Richards et al.\
\cite{ric1999}) contains only 10 other quasars with a greater number
of absorbers. Of these quasars, only two (Q0958+551 and Q1225+317) are
of comparable brightness, but both are at considerably lower
redshifts.

HS\,1603+3820 is even more spectacular when associated absorption
spectrum is concerned. York's catalog contains only three quasars
which have eight or more systems within 8000~km\,s$^{-1}$ of the
emission redshift (Q1037--270, Q1511+091 and Q1556+335). All three are
much fainter; among QSOs brighter than 16~mag no object comes close to
HS\,1603+3820.

We can hypothesize that since none of the absorbers in the complex
appears to be drastically different than the other ones they all may
have been produced by a single event in the quasar's past, and that
they may represent the velocity dispersion of shells of matter ejected
during this event. On the other hand, the spectrum also contains an
absorber which is much stronger than the systems from the complex, as
well as two infalling absorbers, which indicate that the environment
of HS\,1603+3820 is more complicated.

We stress that this paper presents an analysis based on the
observations in relatively low resolution. Our conclusions are by
necessity mostly qualitative since the resolution is inadequate for
performing detailed studies of chemical composition or velocity
structure of the individual systems. High resolution studies of the
complex will be of special interest, since it is heavily blended in
our data and it is very likely that high resolution spectrum will
reveal more absorption systems. HS\,1603+3820 is very bright for a
$z=2.51$ quasar and is therefore an excellent candidate for such
observations.


\acknowledgements

We would like to thank T.~Aldcroft, J.~Bechtold, D.~Dobrzycka,
M.~Elvis, S.~Mathur, J.~Scott, and A.~Siemiginowska for helpful
discussions, A.~Milone for assistance at the MMT, and F.~Drake for
providing good working environment for this project. T.~Aldcroft and
J.~Bechtold wrote computer codes that were used in the analysis.
A.D. acknowledges support from NASA Contract No.~NAS8-39073 (Chandra
X-Ray Observatory Center). The Hamburg Quasar Survey is supported by
the DFG through grants Re~353/11 and Re~353/22.


\end{document}